\documentstyle[11pt]{article}
\input amssymb.sty

\title{Quantum Mechanics and the Interpretation\\of the Orthomodular Square of Opposition}

\author{{\sc C. de Ronde}\thanks{Fellow of the Consejo
Nacional de Investigaciones Cient\'{\i}ficas y T\'ecnicas (CONICET)
} \ $^{1,2}$, {\sc H. Freytes}$^{*}$
$^{3,4}$ and {\sc G. Domenech}\ $^{1,5}$}
\date{}

\begin{document}

\bibliographystyle{plain}
\maketitle

\begin{center}
\begin{small}
1. Center Leo Apostel (CLEA)\\
2. Foundations of  the Exact Sciences (FUND) \\
Brussels Free University  Krijgskundestraat 33, 1160 Brussels -
Belgium \\
3. Universita degli Studi di Cagliari, Via Is Mirrionis 1, 09123,
Cagliari - Italia \\
4. Instituto Argentino de Matem\'atica (IAM) \\
Saavedra 15 - 3er piso - 1083 Buenos Aires, Argentina
5. Instituto de Astronom\'{\i}a y F\'{\i}sica del Espacio (IAFE)\\
Casilla de Correo 67, Sucursal 28, 1428 Buenos Aires - Argentina\\
\end{small}
\end{center}

\begin{abstract}

\noindent In this paper we analyze and discuss the historical and philosophical development of the notion of logical possibility focusing on its specific meaning in classical and quantum mechanics. Taking into account the logical structure of quantum theory we continue our discussion regarding the Aristotelian Square of Opposition in orthomodular structures enriched with a monadic quantifier \cite{FRD12}. Finally, we provide an interpretation of the {\it Orthomodular Square of Opposition} exposing the fact that classical possibility and quantum possibility behave formally in radically different manners.

\end{abstract}

\begin{small}

{\em Keywords: Actuality, Potentiality, Physical Property, Quantum
Logic.}

{\em PACS numbers: 02.10 De}

\end{small}

\bibliography{pom}

\begin{thebibliography}{10}

\bibitem{Aristotle} Aristotle, 1995, {\it The Complete Works of Aristotle}, The Revised Oxford
Translation, J. Barnes (Ed.), Princeton University Press,
New Jersey.

\bibitem{Dieks10} Dieks, D., 2010, ``Quantum Mechanics, Chance and
Modality", {\it Philosophica}, {\bf 82}, forthcoming.

\bibitem{DF} Domenech, G. and Freytes, H., 2005, ``Contextual
logic for quantum systems'', {\it Journal of Mathematical Physics},
{\bf 46}, 012102-1 - 012102-9.

\bibitem{DFR06} Domenech, G., Freytes, H. and de Ronde, C., 2006,
``Scopes and limits of modality in quantum mechanics",
\textit{Annalen der Physik}, {\bf 15}, 853-860.

\bibitem{DFR09} Domenech, G., Freytes, H. and de Ronde, C., 2009,
``Modal-type orthomodular logic'', {\it Mathematical Logic
Quarterly}, {\bf 3}, 307-319.

\bibitem{FeynmanHibbs65} Feynman, R. P. and Hibbs, A. R., 1965,
{\it Quantum Mechanics and Path Integrals}, McGraw-Hill, New York.

\bibitem{FRD12}  Freytes, H., de Ronde, C. and Domenech, G., 2012,  
``The square of opposition in orthodmodular logic'' In {\it Around and Beyond the Square of Opposition: Studies in Universal Logic}, pp. 193-201, Jean-Yves BŽziau and Dale Jacquette (Eds.), Springer, Basel.

\bibitem{HAL}  Halmos, P., 1955, ``Algebraic logic I, monadic Boolean algebras'', {\it Compositio Mathematics},  {\bf 12}, 217-249.

\bibitem{Heis58} Heisenberg, W., 1958, {\it Physics and Philosophy},
World perspectives, George Allen and Unwin Ltd., London.

\bibitem{Ka}  Kalman, J. A., 1958, ``Lattices with involution'', {\it Transactions of the American Mathematical Society}  {\bf 87}, 485-491.

\bibitem{KAL}  Kalmbach, G., 1983,  {\it Ortomodular Lattices}, Academic Press, London.

\bibitem{KarakostasHadzidaki05} Karakostas, V. and Hadzidaki, P.,
2005, ``Realism vs. Constructivism in Contemporary Physics: The
Impact of the Debate on the Understanding of Quantum Theory and its
Instructional Process'', {\it Science $\&$ Education}, {\bf 14},
607-629.

\bibitem{KS} Kochen, S. and Specker, E., 1967, ``On the problem
of Hidden Variables in Quantum Mechanics", {\it Journal of
Mathematics and Mechanics}, {\bf 17}, 59-87. Reprinted in Hooker,
1975, 293-328.

\bibitem{MM} Maeda, F. and  Maeda, S., 1970,  {\it Theory
of Symmetric Lattices}, Springer-Verlag, Berlin.

\bibitem{PauliJung} Pauli, W. and Jung, C. G., 2001, {\it Atom and
Archetype, The Pauli/Jung Letters 1932-1958}, Princeton University
Press, New Jersey.

\bibitem{deRonde10} de Ronde, C., 2010, ``For and Against Metaphysics
in the Modal Interpretation of Quantum Mechanics", {\it
Philosophica}, {\bf 83}, 85-117.

\bibitem{RFD13} de Ronde, C. Freytes, H. and Domenech, G., 2013, ``Interpreting the Modal Kochen-Specker Theorem: Possibility and Many Worlds in Quantum Mechanics'', preprint.

\bibitem{Sambursky88} Sambursky, S., 1988,  {\it The Physical World of the
Greeks}, Princeton University Press, Princeton.

\bibitem{Smets05} Smets, S., 2005,  ``The Modes of Physical Properties
in the Logical Foundations of Physics'', {\it Logic and Logical
Philosophy}, {\bf 14}, 37-53.

\bibitem{VerelstCoecke} Verelst, K. and Coecke, B., 1999, ``Early
Greek Thought and perspectives for the Interpretation of Quantum
Mechanics: Preliminaries to an Ontological Approach'', In {\it The
Blue Book of Einstein Meets Magritte}, 163-196, D. Aerts (Ed.),
Kluwer Academic Publishers, Dordrecht.


\end{thebibliography}

\newtheorem{theo}{Theorem}[section]

\newtheorem{definition}[theo]{Definition}

\newtheorem{lem}[theo]{Lemma}

\newtheorem{met}[theo]{Method}

\newtheorem{prop}[theo]{Proposition}

\newtheorem{coro}[theo]{Corollary}

\newtheorem{exam}[theo]{Example}

\newtheorem{rema}[theo]{Remark}{\hspace*{4mm}}

\newtheorem{example}[theo]{Example}

\newcommand{\proof}{\noindent {\em Proof:\/}{\hspace*{4mm}}}

\newcommand{\qed}{\hfill$\Box$}

\newcommand{\ninv}{\mathord{\sim}} 

\section{The Modes of Physical Properties: Actuality and Potentiality}

The debate in Pre-Soctratic philosophy is traditionally understood
as the contraposition of the Heraclitean and the Eleatic schools of
thought \cite{Sambursky88}. Heraclitus was considered as defending
the theory of flux, a doctrine of permanent motion, change and unstability
in the world. This doctrine precluded,
as both Plato and Aristotle stressed repeatedly, the impossibility
to develop certain knowledge about the world.
``This is so because Being, over a lapse of time, has no stability.
Everything that it is at this moment changes at the same time,
therefore it is not. This coming together of Being and non-Being at
one instant is known as the principle of coincidence of opposites.''
\cite[p. 2]{VerelstCoecke} In contraposition to the Heraclitean
school we find Parmenides as the  main character of the Eleatic
school. Parmenides, as interpreted also by Plato and Aristotle,
taught the non-existence of motion and change in reality, reality
being absolutely One, and being absolutely Being. In his famous poem
Parmenides stated maybe the earliest intuitive exposition of the
{\it principle of non-contradiction}; i.e. that which {\it is} can
only {\it be}, that which {\it is not, cannot be}. In order to dissolve the problem of movement, Aristotle
developed a metaphysical scheme in which, through the notions of
{\it actuality} and {\it potentiality}, he was able to articulate
both the Heraclitean and the Eleatic school \cite{Aristotle}. On the one hand, potentiality contained the undetermined, contradictory and non-individual realm of existence, on the other, the mode of being of actuality was determined through the logical principles  of {\it existence} and {\it non-contradiction}; it was through these same principles together with the principle of {\it identity} that the concept of entity was put forward. Through these principles the notion of entity is capable of unifying, of totalizing in terms of a
``sameness'', creating certain stability for knowledge to be possible. This representation or transcendent description of the world is considered by many the origin of metaphysical thought. Actuality is then linked directly to metaphysical representation and understood as characterizing a mode of existence independent of observation. This is the way through which metaphysical thought was able to go beyond the  {\it hic et nunc}, creating a world beyond the world, a world of concepts.

Although Aristotle presents at first both actual and potential realms as ontologically equivalent, from chapter 6 of book $\Theta$, he seems to place actuality in the central axis of his architectonic, relegating potentiality to a mere supplementary role. ``We have distinguished the various
senses of `prior', and it is clear that actuality is prior to
potentiality. [...] For the action is the end, and the actuality is
the action. Therefore even the word `actuality' is derived from
`action', and points to the fulfillment.'' [1050a17-1050a23]
Aristotle then continues to provide arguments in this line which
show ``[t]hat the good actuality is better and more valuable than the
good potentiality.'' [1051a4-1051a17] But, quite independently of the Aristotelian
metaphysical scheme, it could be argued that the idea of potentiality
could be developed in order to provide a mode of existence
independent of actuality. As we shall see in the following, after modern science discarded almost completely the potential realm, it was quantum theory through some of its authors that potentiality became again a key concept for physics. Wolfgang Pauli had seen this path in
relation to the development of quantum mechanics itself. As noted in
a letter to C. G. Jung dated 27 February 1953:

\begin{quotation}
{\small ``Science today has now, I believe, arrived at a stage where
it can proceed (albeit in a way as yet not at all clear) along the
path laid down by Aristotle. The complementarity characteristics of
the electron (and the atom) (wave and particle) are in fact
`potential being,' but one of them is always `actual nonbeing.' That is why one can say that science, being no longer
classical, is for the first time a genuine theory of becoming and no
longer Platonic.'' \cite[p. 93]{PauliJung}}
\end{quotation}

But before arriving to QM let us first analyze the relation between classical physics and the hilemorphic tradition.

\section{The Actual Realm and Classical Physics}

The importance of potentiality, which was first placed by Aristotle
in equal footing to actuality as a mode of existence, was soon
diminished in the history of western thought. As we have seen above,
it could be argued that the seed of this move was already present in
the Aristotelian architectonic, whose focus was clearly placed in
the actual realm. The realm of potentiality, as a different
(ontological) mode of the being was neglected becoming not more than
mere (logical) {\it possibility}, a process of
fulfillment. In relation to the development of physics, the
focus and preeminence was also given to actuality. The XVII century
division between `res cogitans' and `res extensa' played in this
respect an important role separating very clearly the realms of actuality
and potentiality. The philosophy which was developed after Descartes
kept `res cogitans' (thought) and `res extensa' (entities as
acquired by the senses) as separated realms.\footnote{While `res
cogitans', the soul, was related to the {\it indefinite} realm of
potentiality, `res extensa', i.e. the entities as characterized by the principles of logic, related to the actual.}

\begin{quotation}
{\small ``Descartes knew the undisputable necessity of the
connection, but philosophy and natural science in the following
period developed on the basis of the polarity between the `res
cogitans' and the `res extensa', and natural science concentrated
its interest on the `res extensa'. The influence of the Cartesian
division on human thought in the following centuries can hardly be
overestimated, but it is just this division which we have to
criticize later from the development of physics in our
time.'' \cite[p. 73]{Heis58}}
\end{quotation}

\noindent This materialistic conception of science based itself on
the main idea that extended things exist as being definite, that is,
in the actual realm of existence. With modern science the actualist
Megarian path was recovered and potentiality dismissed as a
problematic and unwanted guest. The transformation from medieval to
modern science coincides with the abolition of Aristotelian
hilemorphic metaphysical scheme ---in terms of potentiality and
actuality--- as the foundation of knowledge. However, the basic
structure of his metaphysical scheme and his logic still remained
the basis for correct reasoning. As noted by Verelst and Coecke:

\begin{quotation}
{\small ``Dropping Aristotelian metaphysics, while at the same time
continuing to use Aristotelian logic as an empty `reasoning
apparatus' implies therefore loosing the possibility to account for
change and motion in whatever description of the world that is based
on it. The fact that Aristotelian logic transformed during the
twentieth century into different formal, axiomatic logical systems
used in today's philosophy and science doesn't really matter,
because the fundamental principle, and therefore the fundamental
ontology, remained the same ([40], p. xix). This `emptied' logic
actually contains an Eleatic ontology, that allows only for static
descriptions of the world."
\cite[p. 7]{VerelstCoecke}}
\end{quotation}

It was Isaac Newton who was able to translate into a closed
mathematical formalism both, the ontological presuppositions present
in Aristotelian (Eleatic) logic and the materialistic ideal of `res
extensa' together with actuality as its mode of existence. 
In classical mechanics the representation of the state of the
physical system is given by a point in phase space $\Gamma$ and the
physical magnitudes are represented by real functions over $\Gamma$.
These functions commute in between each others and can be
interpreted as possessing definite values independently of
measurement, i.e. each function can be interpreted as being actual.
The term actual refers here to {\it preexistence} (within the
transcendent representation) and not to the observation {\it hic et
nunc}. Every physical system may be described
exclusively by means of its actual properties. The change of the system may be
described by the change of its actual properties. Potential or
possible properties are considered as the points to which the system
might arrive in a future instant of time. As also noted by Dieks:

\begin{quotation}
{\small ``In classical physics
the most fundamental description of a physical system (a point in
phase space) reflects only the actual, and nothing that is merely
possible. It is true that sometimes states involving probabilities
occur in classical physics: think of the probability distributions
$\rho$ in statistical mechanics. But the occurrence of possibilities
in such cases merely reflects our ignorance about what is actual.
The statistical states do not correspond to features of the actual
system (unlike the case of the quantum mechanical superpositions),
but quantify our lack of knowledge of those actual features.''
\cite[p. 124]{Dieks10}}
\end{quotation}

\noindent Classical mechanics tells us via the equation of motion how the
state of the system moves along the curve determined by the initial
conditions in $\Gamma$ and thus, as any mechanical property
may be expressed in terms of $\Gamma$'s variables, how all of them
evolve. Moreover, the structure in which actual
properties may be organized is the (Boolean) algebra of classical
logic.

\section{Heisenberg and the Recovery of the Potential Realm}

The  mechanical description of the world provided by Newton can be
sketched in terms of static pictures which provide at each instant
of time the set of definite actual properties within a given state
of affairs \cite[p. 609]{KarakostasHadzidaki05}. Obviously
there is in this description a big debt to the Aristotelian
metaphysical scheme. However, the description of motion is then
given, not {\it via} the path from the potential to the
actual, from {\it matter} into {\it form}, but rather {\it via} the
successions of actual states of affairs; i.e., stable situations,
``pictures'', constituted by sets of actual properties with definite
values. As we discussed above, potentiality becomes then
superfluous. With the advenment of modern science and the
introduction of mathematical schemes, physics seemed capable of
reproducing the evolution of the universe. The idea of an actual
state of affairs (i.e. the set of actual properties which
characterize a system) supplemented by the dynamics allowed then to
imagine a Demon such as that of Laplace capable of knowing the past
and future states of the universe. If we could know the actual
values at the definite instant of time we could also derive the
actual set of properties in the future and the past. As Heisenberg
explains, this materialistic conception of science chose actuality
as the main aspect of existence:

\begin{quotation}
{\small ``In the philosophy of Aristotle, matter was thought of in
the relation between form and matter. All that we perceive in the
world of phenomena around us is formed matter. Matter is in itself
not a reality but only a possibility, a `potentia'; it exists only
by means of form. In the natural process the `essence,' as Aristotle
calls it, passes over from mere possibility through form into
actuality. [...] Then, much later, starting from the philosophy of
Descartes, matter was primarily thought of as opposed to mind. There
were the two complementary aspects of the world, `matter' and
`mind,' or, as Descartes put it, the `res extensa' and the `res
cogitans.' Since the new methodical principles of natural science,
especially of mechanics, excluded all tracing of corporeal phenomena
back to spiritual forces, matter could be considered as a reality of
its own independent of the mind and of any supernatural powers. The
`matter' of this period is `formed matter,' the process of formation
being interpreted as a causal chain of mechanical interactions; it
has lost its connection with the vegetative soul of Aristotelian
philosophy, and therefore the dualism between matter and form
[potential and actual] is no longer relevant. It is this concept of
matter which constitutes by far the strongest component in our
present use of the word `matter'.'' \cite[p. 129]{Heis58}}
\end{quotation}

As mentioned above, in classical mechanics the mathematical
description of the behavior of a system may be formulated in terms
of the set of actual properties. The same treatment can be applied
to quantum mechanics. However, the different structure of the
physical properties of the system in the new theory determines a
change of nature regarding the meaning of possibility and
potentiality. Quantum mechanics was related to modality since Born's
interpretation of the quantum wave function $\Psi$ in terms of a
density of probability. But it was clear from the very beginning
that this new quantum possibility was something completely different
from that considered in classical theories. ``[The] concept of the
probability wave [in quantum mechanics] was something entirely new
in theoretical physics since Newton. Probability in mathematics or
in statistical mechanics means a statement about our degree of
knowledge of the actual situation. In throwing dice we do not know
the fine details of the motion of our hands which determine the fall
of the dice and therefore we say that the probability for throwing a
special number is just one in six. The probability wave function,
however, meant more than that; it meant a tendency for something.''
\cite[p. 42]{Heis58} It was Heisenberg who went a step further and
tried to interpret the wave function in terms of the Aristotelian
notion of potentia. Heisenberg argued that the concept of
probability wave ``was a quantitative version of the old concept of
`potentia' in Aristotelian philosophy. It introduced something
standing in the middle between the idea of an event and the actual
event, a strange kind of physical reality just in the middle between
possibility and reality.'' According to Heisenberg, the concept of
potentiality as a mode of existence has been used implicitly or
explicitly in the development of quantum mechanics:

\begin{quotation}
{\small ``I believe that the language actually used by physicists
when they speak about atomic events produces in their minds similar
notions as the concept of `potentia'. So physicists have gradually
become accustomed to considering the electronic orbits, etc., not as
reality but rather as a kind of `potentia'.'' \cite[p. 156]{Heis58}}
\end{quotation}

\noindent In this respect, one of the most interesting examples of
an implicit use of these ideas has been provided by Richard Feynmann
in his path integral approach \cite{FeynmanHibbs65}. Even though
Feynman talks about calculating probabilities, he seems to refer
implicitly to of existent potentialities. Why, if not, should we
take into account the mutually incompatible paths of the electron in
the double-slit experiment? His approach considers every path as
existent in the mode of being of potentiality, there where the
constrains of actuality cannot be applied. But as we discussed
elsewhere \cite{deRonde10}, Heisenberg's attempt to interpret
quantum mechanics with a non-classical conceptual scheme might have
been highly compromised by Bohr's own agenda. In any case, we must
admit that apart from some few remarks and analogies, Heisenberg's
interpretation remained not only incomplete but also unclear in many
aspects.

\section{Quantum Possibility in the Orthomodular Structure}

Elsewhere we have discussed the importance of distinguishing, both from a formal and conceptual level the notion of (classical) possibility ---arising in the distributive Boolean structure--- from that of quantum possibility ---arising from an orthomodular structure. In order to discuss some interpretational aspects of quantum possibility we first recall from \cite{KAL, MM} some notions about orthomodular lattices.  A {\it lattice with involution} \cite{Ka} is an algebra $\langle \mathcal{L}, \lor, \land, \neg \rangle$ such that $\langle
\mathcal{L}, \lor, \land \rangle$ is a  lattice and $\neg$ is a
unary operation on $\mathcal{L}$ that fulfills the following
conditions: $\neg \neg x = x$ and $\neg (x \lor y) = \neg x \land
\neg y$.  An {\it orthomodular lattice} is an algebra $\langle {\cal L},
\land, \lor, \neg, 0,1 \rangle$   of type $\langle
2,2,1,0,0 \rangle$ that satisfies the following conditions:

\begin{enumerate}
\item
$\langle {\cal L}, \land, \lor, \neg, 0,1 \rangle$  is a bounded
lattice with involution,

\item
$x\land  \neg x = 0 $.

\item
$x\lor ( \neg x \land (x\lor y)) = x\lor y $

\end{enumerate}

We denote by ${\cal OML}$ the variety of orthomodular lattices. Let
$\mathcal{L}$ be an orthomodular lattice. {\it Boolean algebras} are
orthomodular lattices satisfying  the {\it distributive law} $x\land
(y \lor z) = (x \land y) \lor (x \land z)$. We denote by ${\bf 2}$
the Boolean algebra of two elements. Let $\mathcal{L}$ be an
orthomodular lattice. An element $c\in \mathcal{L}$ is said to be a
{\it complement} of $a$ iff $a\land c = 0$ and $a\lor c = 1$. Given
$a, b, c$ in $\mathcal{L}$, we write: $(a,b,c)D$\ \   iff $(a\lor
b)\land c = (a\land c)\lor (b\land c)$; $(a,b,c)D^{*}$ iff $(a\land
b)\lor c = (a\lor c)\land (b\lor c)$ and $(a,b,c)T$\ \ iff
$(a,b,c)D$, (a,b,c)$D^{*}$ hold for all permutations of $a, b, c$.
An element $z$ of $\mathcal{L}$ is called {\it central} iff for all
elements $a,b\in L$ we have\ $(a,b,z)T$. We denote by
$Z(\mathcal{L})$ the set of all central elements of $\mathcal{L}$
and it is called the {\it center} of $\mathcal{L}$.

\begin{prop}\label{eqcentro} Let $\mathcal{L}$ be an orthomodular lattice. Then we have:

\begin{enumerate}

\item
$Z(\mathcal{L})$ is a Boolean sublattice of $\mathcal{L}$ {\rm
\cite[Theorem 4.15]{MM}}.

\item
$z \in Z(\mathcal{L})$ iff for each $a\in \mathcal{L}$, $a = (a\land
z) \lor (a \land \neg z)$  {\rm \cite[Lemma 29.9]{MM}}.

\end{enumerate}
\qed
\end{prop}

In the orthodox formulation of quantum mechanics, a property of (or
a proposition about) a quantum system is related to a closed
subspace of the Hilbert space ${\mathcal H}$ of its (pure) states
or, analogously, to the projector operator onto that subspace. Physical properties of the system are organized in the orthomodular lattice of closed
subspaces ${\mathcal L}({\mathcal H})$ also called {\it Hilbert lattice}.  
Let ${\mathcal H}$ be Hilbert space representing a quantum system. Differently from the classical scheme, a physical magnitude ${\mathcal M}$ is represented by an operator ${\bf M}$
acting over the state space. For bounded self-adjoint operators,
conditions for the existence of the spectral decomposition ${\bf
M}=\sum_{i} a_i {\bf P}_i=\sum_{i} a_i |a_i\rangle\langle a_i|$ are
satisfied. The real numbers $a_i$ are related to the outcomes of
measurements of the magnitude ${\mathcal M}$ and projectors
$|a_i\rangle\langle a_i|$ to the mentioned properties. Each self-adjoint operator
$\bf M$ has associated a Boolean sublattice $W_{\bf{M}}$ of
${\mathcal L}({\mathcal H})$ which we will refer to as the spectral
algebra of the operator $\bf M$. Assigning values to a physical
quantity ${\cal M}$ is equivalent to establishing a Boolean
homomorphism $v: W_{\bf{M}} \rightarrow {\bf 2}$.

The fact that physical magnitudes are represented by operators on ${\cal H}$ that, in general, do not commute has extremely problematic interpretational consequences for it is then
difficult to affirm that these quantum magnitudes are \emph{simultaneously preexistent}. In order to restrict the discourse to  sets of commuting magnitudes, different Complete Sets
of Commuting Operators (CSCO) have to be chosen. This choice has not found until today a clear justification and remains problematic. This feature is called in the literature {\it quantum contextuality}.
The Kochen-Specker theorem (KS theorem for short) rules out the non-contextual assignment of
definite values to the physical properties of a quantum system \cite{KS}. This
may be expressed in terms of valuations over  ${\mathcal
L}({\mathcal H})$ in the following manner. We first introduce the
concept of global valuation. Let  $(W_i)_{i\in I}$ be the family of
Boolean sublattices of ${\mathcal L}({\mathcal H})$. Then a {\it
global valuation} of the physical magnitudes over ${\mathcal
L}({\mathcal H})$ is a family of Boolean homomorphisms $(v_i: W_i
\rightarrow {\bf 2})_{i\in I}$ such that $v_i\mid W_i \cap W_j =
v_j\mid W_i \cap W_j$ for each $i,j \in I$. If this global valuation
existed, it would allow to give values to all magnitudes at the same
time maintaining a {\it compatibility condition} in the sense that
whenever two magnitudes shear one or more projectors, the values
assigned to those projectors are the same from every context. The KS
theorem, in the algebraic terms, rules out the existence of global
valuations when $dim({\mathcal H})>2$ {\rm \cite[Theorem 3.2]{DF}}.
Contextuality can be directly related to the impossibility to
represent a piece of the world as constituted by a set of definite
valued properties independently of the choice of the context. This
definition makes reference only to the actual realm. But as we know,
QM makes probabilistic assertions about measurement results.
Therefore, it seems natural to assume that QM does not only deal
with actualities but also with possibilities. 

Following \cite{DFR06} we delineate a modal extension for orthomodular
lattices that allows to formally represent, within the same
algebraic structure, actual and possible properties of the system.
This allows us to  discuss the restrictions posed by the theory
itself to the {\it actualization} of  possible properties. Given a
proposition about the system, it is possible to define a context
from which one can predicate with certainty about it together with a
set of propositions that are compatible with it and, at the same
time, predicate probabilities about the other ones (Born rule). In
other words, one may predicate truth or falsity of all possibilities
at the same time, i.e., possibilities allow an interpretation in a
Boolean algebra, i.e., if we refer with $\Diamond P$  to the possibility of $P$
then, $\Diamond P \in Z({\cal L})$. This interpretation of possibility in terms of the Boolean algebra
of central elements of ${\cal L}$ reflects the fact that one can
simultaneously predicate about all possibilities because Boolean
homomorphisms of the form $v:Z({\cal L}) \rightarrow {\bf 2}$ can be
always established. If $P$ is a proposition about the system and $P$
occurs, then it is trivially possible that $P$ occurs. This is
expressed as $P \leq \Diamond P$. Classical consequences that are
compatible with a given property, for example probability
assignments to the actuality of other propositions, shear the
classical frame. These consequences are the same ones as those which
would be obtained by considering the original actual property as a
possible property. This is interpreted as, if $P$ is a property of
the system, $\Diamond P$ is the smallest central element greater
than $P$, i.e. $\Diamond P = Min \{z\in Z({\mathcal{L}}): P\leq z \}$.
This enriched orthomodular structure called {\it Boolean 
saturated orthomodular lattices} can be axiomatized by
equations conforming a variety denoted by ${\cal OML}^\Diamond$ \cite[Theorem
4.5]{DFR06}. Orthomodular complete lattices are examples
of Boolean saturated orthomodular lattices. We can embed each orthomodular lattice $\mathcal{L}$ in an element $\mathcal{L}^{\Diamond} \in  {\cal OML}^\Diamond$ {see \rm \cite[Theorem 10]{DFR06}}. 
The {\it modal extension} of $\mathcal{L}$, namely $\mathcal{L}^{\Diamond}$, represents the fact that each orthomodular system can be modally enriched in such a way as to obtain a new propositional system that includes the original propositions in addition to their possibilities. Let $\mathcal{L}$ be an orthomodular lattice and $\mathcal{L}^{\Diamond}$ a modal extension of $\mathcal{L}$. We define the possibility space of $\mathcal{L}$ in $\mathcal{L}^{\Diamond}$ as as the subalgebra of $\mathcal{L}^{\Diamond}$ generated by the set $\{\Diamond (P): P \in {\cal L} \}$. This algebra is denoted by $\Diamond {\cal L}$ and we can prove that  it is a Boolean subalgebra of the modal extension.
Even though the modal extension $\mathcal{L}^{\Diamond}$ of $\mathcal{L}$ represents the complete propositional system. The possibility space represents a classical structure in which only the possibilities added to the discourse about properties of the system are organized.  Within this frame, the actualization of a possible property acquires a rigorous meaning. Let ${\cal L}$ be an orthomodular lattice,
$(W_i)_{i \in I}$ the family of Boolean sublattices of ${\cal L}$
and ${\cal L}^\Diamond$ a modal extension of $\cal L$. If $f:
\Diamond {\cal L} \rightarrow {\bf 2}$ is a Boolean homomorphism, an
actualization compatible with $f$ is a global valuation $(v_i: W_i
\rightarrow {\bf 2})_{i\in I}$ such that $v_i\mid W_i \cap \Diamond
{\cal L} = f\mid W_i \cap \Diamond {\cal L} $ for each $i\in I$.
{\it Compatible actualizations} represent the passage from
possibility to actuality, they may be regarded as formal constrains
when applying the interpretational rules proposed in the different
modal versions. When taking into account compatible actualizations
from different contexts, an analogous of the KS theorem --- which we
have called Modal Kochen Specker (MKS) for obvious reasons--- holds
for possible properties.

\begin{theo}\label{ksm} {\rm \cite[Theorem 6.2]{DFR06}}
Let $\cal L$ be an orthomodular lattice. Then $\cal L$ admits a
global valuation iff for each possibility space there exists a
Boolean homomorphism  $f: \Diamond {\cal L} \rightarrow {\bf 2}$
that admits  a compatible actualization.\qed
\end{theo}

\noindent The MKS theorem shows that no enrichment of the
orthomodular lattice with modal propositions allows to circumvent
the contextual character of the quantum language. Thus, from a
formal perspective, one is forced to conclude that quantum
possibility is something different from classical possibility. 

The larger structure allows to compare the classical and quantum
cases. In the classical case, the elements $A \in \wp (\Gamma )$
interpreted as the properties of the system are part of a Boolean
algebra (with $\Gamma$ the classical phase space and $\wp (\Gamma)$
its power set). The elements of the corresponding modal structure
are constructed by applying the possibility operator $\Diamond$ to
the elements $A$. These new elements $\Diamond A$, that belong to
the modal structure, correspond to possible properties as spoken in
the natural language. However, in this case, the seemingly larger
structure that includes both actual and modal propositions does not
enlarge the expressive power of the language. This is due to the
fact that there exists a trivial correspondence between any pair of
classical  valuations $v_{c}$ and $w_c$  of the actual and the
possible structures to truth-falsity. This relation can be written
as follows: let $A_k\in\wp(\Gamma)$, $k$ a fix index, then
$$w_c(\Diamond A_k) = 1 \Leftrightarrow v_c(A_k) = 1$$
$$w_c(\Diamond A_k) = 0 \Leftrightarrow v_c(A_k) = 0$$
\noindent Thus, given the state of a classical system, possible properties at
a certain time coincide with (simultaneous) actual ones, they may be
identified. And the distinction between the two sets of properties
is never made. In fact, when referring to possible properties of a
system in a given state, one is always making reference to
\emph{future} possible values of the magnitudes, values that are
determined because they are the evaluation of functions at points
$(p, \ q)$ in $\Gamma$ at  future times. These points are
determined by the equation of motion. Thus, not even future
possibilities are classically indeterminate and they coincide with
\emph{future actual properties}. On the contrary, in the quantum
case, the projectors ${\bf P}_a=|a\rangle\langle a|$ on $\mathcal{H}
$, which are interpreted as the properties of a system, belong to an
orthomodular structure. As we have mentioned above, the orthomodular
lattice is enlarged with its modal content by adding the elements
$\Diamond_{Q} |a\rangle\langle a|$. Due to the fact that there is no
trivial relation between the valuations of subsets of the possible
and actual elements to truth-falsity, this new structure  genuinely
enlarges the expressive power of the language. Formally, if
$w_q(\Diamond_{Q} {\bf P}_k) = 1$, with ${\bf P}_k\in W_i$, then
there exists a valuation $v_q$ such that $v_q({\bf P}_k) = 1$ and
another $v'_q$ such that $v'_q({\bf P}_k) = 0$. Thus, contrary to
the classical case, even at the same instant of time, we may
consider two different kind of properties, two different realms,
possible and actual, that do not coincide.

\section{${\cal OML}^\Diamond$-Square of Opposition}

As we have previously discussed, the restriction of the notion of potentiality to that of logical possibility has been of great importance for the development of modern science. The need to interpret QM suggests the reconsideration of this notion in the light of its non-classical structure.  In order to do so, we have studied the Aristotelian square of opposition in  ${\cal OML}^\Diamond$. Such a version of the Square of Opposition is also called {\it Modal Square of Opposition} (MSO) and expresses the essential properties of the monadic first order quantifiers $\forall$, $\exists$. In an algebraic approach, these properties can be represented within the frame of monadic Boolean algebras \cite{HAL}. More precisely, quantifiers are considered as modal operators acting on a Boolean algebra while the MSO is represented by relations between certain terms of the language in which  the algebraic structure is formulated. 

\vspace{0.6cm}

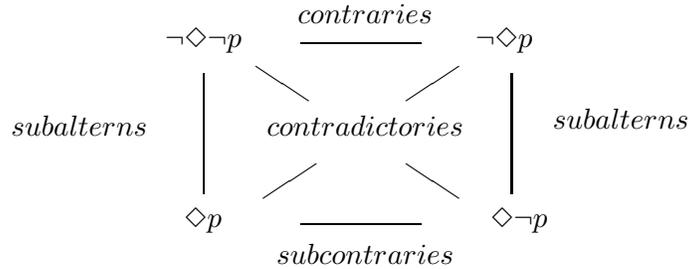
\begin{center}
\unitlength=1mm
\begin{picture}(20,20)(0,0)
\put(3,16){\line(3,0){16}} \put(-10,12){\line(0,-2){16}}
\put(3,-8){\line(1,0){16}} \put(31,12){\line(0,-2){16}}

\put(-10,16){\makebox(0,0){$\neg \Diamond \neg p$}}
\put(30,16){\makebox(0,0){$\neg \Diamond p$}}
\put(-10,-8){\makebox(0,0){$\Diamond p$}}
\put(32,-8){\makebox(0,0){$\Diamond \neg p$}}

\put(4,20){\makebox(15,0){$contraries$}}
\put(-24,5){\makebox(-5,0){$subalterns$}}
\put(14,-13){\makebox(-5,2){$subcontraries$}}
\put(46,5){\makebox(-1,2){$subalterns$}}
\put(12,4){\makebox(-1,2){$contradictories$}}

\put(-3,13){\line(3,-2){7}} \put(17,0){\line(3,-2){7}}

\put(24,13){\line(-3,-2){7}} \put(5,0){\line(-3,-2){7}}

\end{picture}
\end{center}

\vspace{1.5cm}

\noindent The interpretations given to $\Diamond$ from different modal logics determine the corresponding versions of the MSO and by changing the underlying Boolean structure we obtain several generalizations of the monadic first order logic. In what follows we shall interpret this MSO in ${\cal OML}^\Diamond$.  This version of the MSO will be referred as ${\cal OML}^\Diamond$-Square of Opposition. 

Let ${\cal L}$ be an orthomodular lattice and $p \in {\cal L}$ such that $p \not \in Z({\cal L})$, i.e. $p$ can be seen as a non classical proposition in a quantum system represented by ${\cal L}$. Let ${\cal L}^\Diamond$ be a modal extension of ${\cal L}$, $W$ be a Boolean subalgebra of ${\cal L}$, i.e. a context, such that $p\in W$ and consider a classically expanded context $W^\Diamond$ defined as the sub-algebra of ${\cal L}^\Diamond$ generated by $W \cup Z(\mathcal{L}^{\Diamond})$. To analyze the Square, first of all we recall that $\neg p$ is the orthocomplement of $p$. Thus, $\neg$ does not act as a classical negation. But, when applied to possible properties ($\neg \Diamond p$), $\neg$ acts as a classical negation since $\Diamond p$ is a central element.

\begin{itemize}
\item
$\neg \Diamond \neg p \hspace{0.2cm} \underline{contraries} \hspace{0.2cm} \neg \Diamond p $
\end{itemize}

\noindent  Contrary proposition is the negation of the minimum classical consequence of $\neg p$ (the orthogonal complement of $p$) with respect to the negation of the minimum classical consequence of $p$.  
In the usual explanation, two propositions are contrary iff they cannot both be true but can both be false. In our framework we can obtain a similar concept of contrary propositions. Note that $(\neg \Diamond \neg p) \land (\neg \Diamond p) \leq p \land \neg p = 0$. Hence there is not a Boolean valuation $v:W^\Diamond \rightarrow {\bf 2}$ such that $v(\neg \Diamond \neg p) = v(\neg \Diamond p) = 1$, i.e. $\neg \Diamond \neg p$ and $\neg \Diamond p$ ``cannot both be true'' in each possible classically expanded context $W^\Diamond$.
Since $p \not \in Z({\cal L})$, it is not very hard to see that $\Diamond p \land \Diamond \neg p \not = 0$. Then there exists a Boolean valuation $v:W^\Diamond \rightarrow {\bf 2}$ such that 
$v(\Diamond p \land \Diamond \neg p) = 1$. Thus $0 = \neg v(\Diamond p \land \Diamond \neg p) = v(\neg \Diamond p) \lor v(\neg \Diamond \neg p)$. Hence $\neg \Diamond p$ and $\neg \Diamond \neg p$ can both be false.

\begin{itemize}
\item
$\Diamond p \hspace{0.2cm} \underline{subcontraries} \hspace{0.2cm} \Diamond \neg p $
\end{itemize}

\noindent The sub-contrary proposition is the smallest classical consequence
of $p$ with respect to the smallest classical consequence of $\neg
p$. Note that sub-contrary propositions do not depend on the
context. In the usual explanation, two propositions are sub-contrary iff they
cannot both be false but can both be true. Suppose that there exists a Boolean homomorphism $v: W^\Diamond \rightarrow {\bf 2}$ such that $v(\Diamond p) = v(\Diamond \neg p) = 0$. Since $p \leq \Diamond p$ and $\neg p \leq \Diamond \neg p$ then $v(p) = v(\neg p) = 0$ which is a contradiction. Then they cannot both be false. Since $p \not \in Z({\cal L})$, it is not very hard to see that $\Diamond p \land \Diamond \neg p \not = 0$. Hence there exists a Boolean homomorphism $v: W^\Diamond \rightarrow {\bf 2}$ such that $1 = v(\Diamond p \land \Diamond \neg p)= v(\Diamond p) \land v(\Diamond \neg p)$. Then $\Diamond p$ and $\Diamond \neg p$ can both be true.

\begin{itemize}
\item
$\neg \Diamond \neg p \hspace{0.2cm} \underline{subalterns} \hspace{0.2cm} \Diamond p $ and $\neg \Diamond p \hspace{0.2cm} \underline{subalterns} \hspace{0.2cm} \Diamond \neg p $
\end{itemize}

\noindent We study the subalterns propositions $\neg \Diamond \neg p$ and $\Diamond p$  since the other case is analog. For our case, a subaltern proposition is the negation, in the classical sense, of the minimum classical consequence of $\neg p$ (the orthogonal complement of $p$) with respect to the minimum classical consequence of $p$. In the usual explanation, a proposition is subaltern to another one, called {\it superaltern}, iff it must be true when its superaltern
is true and the superaltern must be false when its subaltern is
false. In our case  $\neg \Diamond \neg p$ is superaltern
of $\Diamond p$ and $\neg \Diamond p$ is superaltern
of $\Diamond \neg p$. Since $\neg \Diamond \neg p \leq p \leq \Diamond p$, for
each valuation $v:W^\Diamond \rightarrow {\bf 2}$, if $v(\neg \Diamond \neg p) =
1$ then $v(\Diamond p) = 1$ and if if $v(\Diamond p) = 0$ then
$v(\neg \Diamond \neg p) = 0$ as is required in the subalterns propositions.

\begin{itemize}
\item
$\neg \Diamond \neg p \hspace{0.2cm} \underline{contradictories} \hspace{0.2cm} \Diamond \neg p $ and $\Diamond p \hspace{0.2cm} \underline{contradictories} \hspace{0.2cm} \neg \Diamond p $
\end{itemize}

\noindent The notion of contradictory propositions can be reduced to the relation between  $\Diamond p$ and $\neg \Diamond p$. Contradictory
propositions are the minimum classical consequence of $p$ with respect to the negation of its minimum classical consequence. In the usual explanation, two propositions are contradictory iff they cannot both be true and they cannot both be false. Since $\Diamond p$ is a central element, this property is trivially maintained for $\Diamond p$ and $\neg \Diamond p$.\\

We wish to remark that in terms of valuations the ${\cal OML}^\Diamond$-Square of Opposition behaves in analogous manner to the traditional Square of Opposition, the essential difference being that the concept of ``opposed proposition" represented in the ${\cal OML}^\Diamond$-Square of Opposition is rooted in the concept of minimum classical consequence of a property of a physical system. This fact manifests itself in the following: if $p$ were a classic proposition, i.e. $p\in Z({\cal L})$, then $p = \Diamond p$. In other words, the minimum classical consequence of $p$ is itself. In this way the concept of minimum classical consequence is trivialized, and consequently, also the Square. More precisely, on the one hand the concepts of contradictories, contraries and subcontraries propositions collapse with the classical contradiction $\{p, \neg p\}$ and the subalternation collapses to the trivial equivalence $p\leftrightarrow p$.\\           

The previous analysis exposes once again the fact that classical possibility and quantum possibility formally behave in different manners. This argument adds to the discussion provided in \cite{RFD13} calling the attention to the misinterpretation of the notion of possibility in QM.

\section*{Acknowledgements}

This work was partially supported by the following grants: PIP 112-201101-00636, Ubacyt 2011/2014 635, FWO project G.0405.08 and FWO-research community W0.030.06. CONICET RES. 4541-12 (2013-2014).

\end{document}